\documentclass[aps,twocolumn,showpacs,superscriptaddress,amsmath,amssymb,prl,nofootinbib]{revtex4-1}
\usepackage{epsf,amsmath,amssymb,amsfonts,verbatim,color,multirow,pifont}
\usepackage{graphicx}
\usepackage{dcolumn}
\usepackage{bm}
\usepackage{txfonts}
\usepackage{hyperref}
\usepackage{soul}

\begin{document}
\title{Brownian heat engine with active reservoirs}
\author{Jae Sung Lee}
\author{Jong-Min Park}
\author{Hyunggyu~Park}
\email{hgpark@kias.re.kr}
\affiliation{School of Physics and Quantum Universe Center, Korea Institute for Advanced Study, Seoul 02455, Korea}

\newcommand{\revise}[1]{{\color{red}#1}}

\date{\today}

\begin{abstract}
Microorganisms such as bacteria are active matters which consume chemical energy and generate their unique run-and-tumble motion. A swarm of such microorganisms provide a nonequilibrium active environment whose noise characteristics are different from those of thermal equilibrium reservoirs. One important difference  is a finite persistence time, which is considerably large compared to that of the equilibrium noise, that is, the active noise is colored. Here, we study a mesoscopic energy-harvesting device (engine) with active reservoirs harnessing this noise nature. For a simple linear model, we analytically show that the engine efficiency can surpass the conventional Carnot bound, thus the power-efficiency tradeoff constraint is released,
and the efficiency at the maximum power can overcome the Curzon-Ahlborn efficiency. We find that the supremacy of the active engine critically depends on the time-scale symmetry of two active reservoirs.
\end{abstract}

\pacs{05.70.Ln, 05.70.-a, 05.60.Gg}

\maketitle

\emph{Introduction} -- Mounting social need on sustainable development has attracted  a great attention on energy harvesting techniques, by which useful energy is extracted from surrounding environments, in both scientific and engineering societies~\cite{energy_harvest_review1,energy_harvest_review2, graphene}. Typical examples are thermoelectric devices using a temperature gradient~\cite{harvesting_thermoelectric}, photovoltaic devices using sunlights~\cite{harvesting_photo}, and piezoelectric devices using ambient pressure~\cite{harvesting_piezo}. A major challenging issue on these studies is achieving a high efficiency as well as a high energy or power production. When a device works in an equilibrium environment, the efficiency is bound by the thermodynamic second law; for example, the efficiency of thermoelectric devices cannot surpass the Carnot efficiency.

Then, how is the efficiency affected by replacing the environment with nonequilibrium reservoirs? One might think that the efficiency would be reduced with nonequilibrium reservoirs as the efficiency usually diminishes with irreversibility. However, this is not always true: It was already reported that the efficiency of a quantum heat engine can surpass the conventional Carnot limit with nonequilibrium {\em squeezed} reservoirs~\cite{squeezed1,squeezed2,squeezed3}. In classical systems, it was experimentally shown that the efficiency of a Stirling engine working in a bacterial bath can overcome its maximum efficiency obtained by a quasistatic operation in equilibrium reservoirs~\cite{bacterial_bath_exp,wijland}. In addition, there are also a few examples where the efficiency increases with the irreversibility in well-manipulated ways~\cite{JSLee1,JSLee2,PoEs}.
However, a systematic study on the efficiency bound of engines working in nonequilibrium environments has rarely been done, partly because its theoretical manipulation  is not straightforward as in the equilibrium cases.

In this work, we study the efficiency and the power of an energy-harvesting device extracting energy from nonequilibrium {\em active} reservoirs. To be specific, we consider an overdamped Brownian motion of passive particles composing of the engine with equilibrium baths and/or bacterial active baths. A bacterial bath is known to be well described by the colored noise
with a finite persistence time scale~\cite{bacteria_exp1,bacteria_exp2,bacteria_exp3,bacteria_exp4,Bechinger,Dabelow}.
In the case with the active baths, we demonstrate rigorously that
(i) the efficiency can surpass the standard Carnot limit, thus the conventional power-efficiency tradeoff relation~\cite{power-eff-rel1,power-eff-rel2,power-eff-rel3} does not hold and (ii) the efficiency at the maximum power (EMP) can overcome the Curzon-Ahlborn (CA) efficiency~\cite{CAefficiency}. We also find that the supremacy of the active engine is
achieved when the time scales of the two active baths are different from each other.
\\

\emph{Engine with equilibrium reservoirs} -- We first revisit the simple linear Brownian engine model with equilibrium reservoirs in the overdamped limit~\cite{Crisanti,ParkJM}.
Suppose that there are two particles (particle $1$ and $2$), each of which moves in a one-dimensional space and is immersed in a heat reservoir with temperature $T_i$ ($i=1,2$). Their positions are denoted by $x_1$ and $x_2$ and $\Phi = \Phi(x_1, x_2)$ is a given potential. The motions of these particles are described by the following equations:
\begin{align}
\gamma_i \dot{x}_i = -\partial_{x_1} \Phi +f_i^\textrm{nc} + \sqrt{2 k_\textrm{B} \gamma_i T_i} \xi_i~,
\label{eq:main_eq}
\end{align}
where $\gamma_i$ is a dissipation coefficient, $f_i^\textrm{nc}$ is an external nonconservative force, and $k_\textrm{B}$ is the Boltzmann constant, which will be set to $1$ in the following discussion. $\xi_i$ is a Gaussian white noise satisfying $\langle \xi_i (t) \rangle = 0$ and $\langle \xi_i (t) \xi_j (t^\prime) \rangle = \delta_{ij} \delta(t-t^\prime)$.
In this model, the harmonic potential and  the linear nonconservative force are taken for analytic treatments~\cite{ParkJM,Chulan, Pietzonka,Chun} as
\begin{align}
\Phi = \frac{k}{2}(x_1^2 + x_2^2)~,\quad ( f_1^\textrm{nc}, f_2^\textrm{nc} ) = (\epsilon x_2, \delta x_1)~.
\label{eq:linear}
\end{align}
Note that
the Brownian gyrator~\cite{Filliger} has a similar structure, which was experimentally realized recently~\cite{Chiang}.

From Eq.~\eqref{eq:main_eq}, the thermodynamic first law can be written as
\begin{align}
\dot{Q}_i =  \dot{E}_i +\dot{W}_i~, \label{eq:firstLaw}
\end{align}
where $\dot{E}_i =  \dot{x}_i \partial_{x_i} \Phi$ is the rate of the internal energy change of a particle $i$, $\dot{W}_i = -f_i^\textrm{nc} \dot{x}_i$ is the work extraction rate due to the external force $f_i^\textrm{nc}$, and $\dot{Q}_i = \dot{x}_i \circ (-\gamma_i \dot{x}_i + \sqrt{2\gamma_i T_i } \xi_i )$ is the heat current out of the bath $i$ with the Stratonovich multiplication denoted by   $\circ$~\cite{Risken}.
In the steady state, $\langle \dot{Q}_i \rangle_\textrm{s} = \langle \dot{W}_i \rangle_\textrm{s} $ as $\langle  \dot{E}_i \rangle_\textrm{s} =0$, where $\langle \cdots \rangle_\textrm{s}$ denotes the steady-state average. Therefore, $\langle \dot{Q}_1 \rangle_\textrm{s} =  -\epsilon \langle  x_2 \dot{x}_1 \rangle_\textrm{s} $ and $\langle \dot{Q}_2 \rangle_\textrm{s} =  - \delta \langle  x_1 \dot{x}_2 \rangle_\textrm{s} $. The total work rate (power) is $\langle \dot{W} \rangle_\textrm{s} =  \langle \dot{W}_1 \rangle_\textrm{s} + \langle \dot{W}_2 \rangle_\textrm{s} = (\epsilon - \delta ) \langle x_1 \dot{x}_2 \rangle_\textrm{s} $, where the second equality comes from the fact
$\frac{d}{dt}\langle x_1 x_2 \rangle_\textrm{s} = \langle \dot{x}_1 x_2 \rangle_\textrm{s}+ \langle x_1 \dot{x}_2 \rangle_\textrm{s}=0 $.

For $T_1 > T_2$, the efficiency $\eta$ is given by the ratio between $\langle \dot{W} \rangle_\textrm{s}$ and $\langle \dot{Q}_1 \rangle_\textrm{s}$ as
\begin{align}
\eta =  \frac{\langle \dot{W} \rangle_\textrm{s}}{ \langle \dot{Q}_1 \rangle_\textrm{s} } = 1 -\frac{\delta}{\epsilon}~. \label{eq:efficiency}
\end{align}
Requiring the work extraction
$\langle \dot{W} \rangle_\textrm{s} = ( \epsilon - \delta ) \langle x_1 \dot{x}_2 \rangle_\textrm{s} \geq 0$ with
\begin{align}
\langle x_1 \dot{x}_2 \rangle_\textrm{s} =\frac{ T_1 \delta - T_2 \epsilon }{ k( \gamma_1 + \gamma_2)  }~, \label{eq:x1v2_normal}
\end{align}
we find the constraint $T_2/T_1 \le \delta/\epsilon \le 1$, leading to the famous
Carnot bound as
\begin{align}
0\le \eta \leq 1-\frac{T_2}{T_1} \equiv \eta_\textrm{C}~, \label{eq:Carnot_bound}
\end{align}
where $\eta_\textrm{C}$ is the Carnot efficiency. As expected from the power-efficiency tradeoff relation, the power $\langle \dot{W} \rangle_\textrm{s}$ vanishes at $\eta = \eta_\textrm{C}$~\cite{power-eff-rel1,power-eff-rel2,power-eff-rel3}.
In addition, we need the stability condition for the existence of the steady state, which turns out to be
\begin{align}
\delta \epsilon \equiv K  < k^2~. \label{eq:stabilityCond}
\end{align}
Derivations of Eqs.~\eqref{eq:x1v2_normal} and \eqref{eq:stabilityCond} are presented in Supplemental Material (SM) I.A.

Figure~\ref{fig:delta-epsilon_plot}(a) shows the engine area satisfying the above two constraints~\eqref{eq:Carnot_bound} and~\eqref{eq:stabilityCond}
with $\gamma_1=\gamma_2 = 1$,  $k=2$, $T_1=2$, and $T_2=1$.
In Fig.~\ref{fig:delta-epsilon_plot}(b), we plot  the normalized efficiency and power as $\tilde{\eta} \equiv \eta/\eta_\textrm{C}$ and $\tilde{P} \equiv \langle \dot{W} \rangle_\textrm{s}/P_\textrm{eq}^\textrm{max} $ along the line from $\epsilon_1$ to $\epsilon_2$ of Fig.~\ref{fig:delta-epsilon_plot}(a) at fixed $\delta=0.8$, where $P_\textrm{eq}^\textrm{max}$ is the maximum power in equilibrium baths defined as Eq.~\eqref{eq:EMP1}. Note that the solid curves and data points of Fig.~\ref{fig:delta-epsilon_plot} are analytic curves and numerical simulation results, respectively. All numerical data are obtained by integrating the equations of motion of Eq.~\eqref{eq:main_eq} and averaging over $2\times 10^6$ samples in the steady state.

We also calculate the efficiency at the maximum power (EMP) $\eta_\textrm{EMP}$.
Along the curve with fixed $K$ $(0\leq K \leq k^2)$, the local maximum power is obtained at
$\delta_\textrm{m}=\sqrt{T_2/T_1} \epsilon_\textrm{m}$ with  the efficiency $\eta_\textrm{m}$ identical to the Curzon-Ahlborn (CA) efficiency $\eta_\textrm{CA}$~\cite{CAefficiency} and the power given by $\langle \dot{W} \rangle_\textrm{s}^\textrm{m}=KT_1 \eta_\textrm{CA}^2 /k(\gamma_1+\gamma_2)$. The global power maximum is achieved at $K=k^2$, thus we obtain
\begin{align}
\eta_\textrm{EMP} = 1 -\sqrt{\frac{T_2}{T_1}} \equiv \eta_\textrm{CA} ~~~\textrm{and}~~ P_\textrm{eq}^\textrm{max} \equiv \frac{kT_1 \eta_\textrm{CA}^2}{\gamma_1+\gamma_2}~.
\label{eq:EMP1}
\end{align}
Figure~\ref{fig:delta-epsilon_plot}(c) shows the plots of $\tilde{\eta}_\textrm{m} \equiv \eta_\textrm{m} /\eta_\textrm{CA}$ and $\tilde{P}_\textrm{m} \equiv \langle \dot{W} \rangle_\textrm{s}^\textrm{m}/  P_\textrm{eq}^\textrm{max} $ against $K$.
\\

\begin{figure*}
\centering
\includegraphics[width=0.95\linewidth]{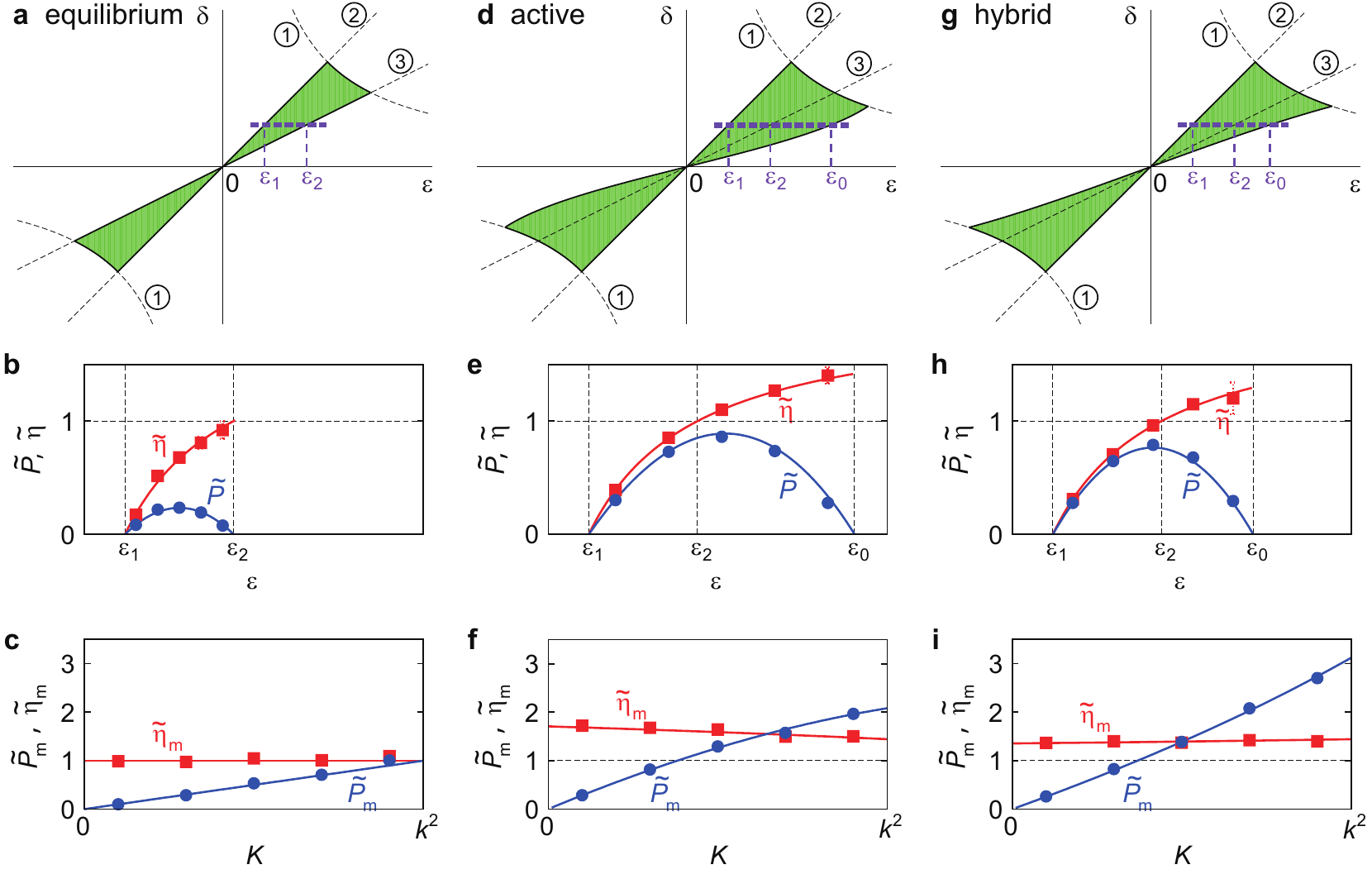}
\caption{Performance of the engine in various environments. Shaded areas in (a), (d), and (g) denote the areas in which the device works as a stable and useful engine for equilibrium, active, and hybrid reservoirs, respectively. \textcircled{\scriptsize 1}, \textcircled{\scriptsize 2}, and \textcircled{\scriptsize 3} denote the curves $\epsilon \delta = k^2$, $\delta=\epsilon$, and $\delta =  \epsilon T_2/T_1$ or $\delta =  \epsilon T_2^\textrm{e}/T_1^\textrm{e}$, respectively. (b), (e), and (h) show the plots of $\tilde{\eta} \equiv \eta/\eta_\textrm{C}$ or $\tilde{\eta} \equiv \eta/\eta_\textrm{C}^\textrm{e}$  and $\tilde{P} \equiv \langle \dot{W} \rangle_\textrm{s}/P_\textrm{eq}^\textrm{max}$ along the {thick dashed} line of (a), (d), and (g) from $\epsilon_1$  to $\epsilon_2$  or $\epsilon_0$ at fixed $\delta =0.8$, respectively.
The efficiency for the active and the hybrid bath can surpass the effective Carnot efficiency $\eta_\textrm{C}^\textrm{e}$ while the efficiency of the equilibrium bath is bounded by the conventional $\eta_\textrm{C}$. Solid curves and data points indicate analytic and numerical results, respectively. (c), (f), and (i) show the plots of $\tilde{\eta}_\textrm{m} \equiv \eta_\textrm{m} /\eta_\textrm{CA}$
or $\tilde{\eta}_\textrm{m} \equiv \eta_\textrm{m} /\eta_\textrm{CA}^\textrm{e}$ and $\tilde{P}_\textrm{m} \equiv \langle \dot{W} \rangle_\textrm{s}^\textrm{m}/P_\textrm{eq}^\textrm{max}$ against $K(=\delta\epsilon)$. The EMP for the active and the hybrid bath surpass $\eta_\textrm{CA}^\textrm{e}$. The maximum power also become larger for the active and the hybrid baths than that for the equilibrium bath.
} \label{fig:delta-epsilon_plot}
\end{figure*}

\emph{Engine with active reservoirs} -- Now, we replace equilibrium reservoirs with bacterial active baths.
The equations of motion are given as
\begin{align}
\gamma_i \dot{x}_i &= -\partial_{x_i} \Phi +f_i^\textrm{nc} + \zeta_i~.
\label{eq:main_eq_colored}
\end{align}
Here, $\zeta_i$ is a Gaussian colored noise satisfying $\langle \zeta_i (t) \zeta_j (t^\prime) \rangle = D_i \delta_{ij} e^{-|t-t^\prime|/\tau_i } /\tau_i$, where $D_i$ is the noise strength and $\tau_i$ is the persistence time scale of noise
$\zeta_i$~\cite{bacteria_exp1,bacteria_exp2,bacteria_exp3,bacteria_exp4,Bechinger,Dabelow}.
The finite persistent time originates from collisions of a passive particle with bacteria with directional persistence.
In the $\tau_i \rightarrow 0$ limit, the active bath becomes identical to the equilibrium bath with the temperature
$T_i = D_i/\gamma_i$.
The Ornstein-Uhlenbeck process (OUP) provides one of the simplest ways to describe the evolution of $\zeta_i$~\cite{Risken}:
\begin{align}
\tau_i \dot{\zeta}_i = - \zeta_i + \sqrt{2 D_i} \xi_i~, \label{eq:main_zetaEq}
\end{align}
where $\xi_i$ is a Gaussian white noise as seen in Eq.~\eqref{eq:main_eq}.
Together with Eq.~\eqref{eq:main_eq_colored}, this process with linear forces like in Eq.~\eqref{eq:linear} is called the active OUP (AOUP)~\cite{Fodor,Madal,Marconi,Dabelow}.
We remark that a non-Gaussian nature of the colored noise was observed experimentally in a low-concentration
bacterial bath~\cite{bacteria_exp3}. However, our results in the following can also apply to a non-Gaussian case because
the work and heat current in the linear-force system do not depend on the higher-order moments of the noise except for the second-order
one~\cite{wijland}.

Before investigating the AOUP engine, we first consider  passive particles trapped in a harmonic potential in Eq.~\eqref{eq:linear} without a nonconservative force ($f_i^\textrm{nc}=0$), in contact with the active reservoir. From its steady state distribution, we can unambiguously define the appropriate {\em effective} temperature of the active reservoir as follows.
It is straightforward (see SM I.B) to derive the steady state distribution which is  Boltzmannlike  in this case as
\begin{align}
P(x_1,x_2) = \prod_{i=1}^2  \sqrt{\frac{k}{2 \pi {T}_i^\textrm{e}  } }  e^{-\frac{k x_i^2 }{2  {T}_i^\textrm{e}}  }, \label{eq:pdf_colored}
\end{align}
where the effective temperature ${T}_i^\textrm{e} \equiv D_i/\Gamma_i$ with $\Gamma_i=\gamma_i+ k\tau_i$.
Note that ${T}_i^\textrm{e} \le D_i/\gamma_i=T_i$ and
depends not only on the persistent time $\tau_i$ but also on the stiffness $k$ of the harmonic potential. It is not surprising to see
the effectively lower temperature because the persistence reduces the stochasticity.

The energy conservation yields again Eq.~\eqref{eq:firstLaw} where the heat current out of the active bath $i$ is given by $\dot{Q}_i = \dot{x}_i \circ(-\gamma_i \dot{x}_i +  \zeta_i )$. In the steady state for the AOUP engine, we get the same form  as before for the heats and power such as $\langle \dot{Q}_1 \rangle_\textrm{s} = \langle \dot{W}_1 \rangle_\textrm{s}= \epsilon \langle  x_1 \dot{x}_2 \rangle_\textrm{s} $ and $\langle \dot{W} \rangle_\textrm{s} =  (\epsilon - \delta ) \langle x_1 \dot{x}_2 \rangle_\textrm{s} $.
The standard calculation of the multivariate OUP~\cite{Risken} by treating the colored
noise $\zeta_i$ as a state variable yields (see SM I.C)
\begin{align}
\langle x_1 \dot{x}_2 \rangle_\textrm{s} =\frac{1}{ k( \gamma_1 + \gamma_2)  }
\left[   \frac{T_1^\textrm{e}\delta}{ \mathcal{A}_{K} }  - \frac{T_2^\textrm{e}\epsilon}{ \mathcal{B}_{K} } \right]\quad \textrm{with}~~K=\delta\epsilon~, \label{eq:x1v2_active}
\end{align}
where
\begin{align}
\mathcal{A}_K &= 1+  \frac{k \gamma_1 \tau_1}{ \gamma_2 \Gamma_1} +  \frac{(k^2 - K)\tau_1^2}{\gamma_2 \Gamma_1}~,\nonumber\\
\mathcal{B}_K &= 1+  \frac{k \gamma_2 \tau_2}{ \gamma_1 \Gamma_2} +  \frac{(k^2 - K)\tau_2^2}{\gamma_1 \Gamma_2}~.
\label{eq:AB_K}
\end{align}
We find the same stability condition ($K<k^2$, see SM I.C), thus $\mathcal{A}_K\ge 1$ and $\mathcal{B}_K\ge 1$ in the stable region.

With the same definition of the efficiency for a passive engine in Eq.~\eqref{eq:efficiency} (see further discussions in SM II), we find the efficiency bound
from the engine condition as
\begin{align}
0\le \eta =1-\frac{\delta}{\epsilon}\leq 1-\frac{T_2^\textrm{e}\mathcal{A}_K}{T_1^\textrm{e}\mathcal{B}_K}\equiv \eta_\textrm{C}^\textrm{a} ~,
\label{eq:new_bound}
\end{align}
where $\eta_\textrm{C}^\textrm{a}$  is the maximum efficiency for the AOUP engine. It is remarkable to see that
$\eta_\textrm{C}^\textrm{a}$ can exceed the effective Carnot efficiency $\eta_\textrm{C}^\textrm{e}=1-{T_2^\textrm{e}}/{T_1^\textrm{e}}$
when the modification factor $\mathcal{M}_{K}\equiv \mathcal{B}_K /\mathcal{A}_K >1$. {The modification factor
$\mathcal{M}_{K}$ is maximized and reaches $1+k\tau_2/\gamma_1$ in the limits of $\tau_1/\tau_2 \rightarrow 0$ and $K\rightarrow 0$. }

The case with the time-scale symmetry ($\tau_1=\tau_2$ and $\gamma_1=\gamma_2$) is special. We get
$\mathcal{M}_{K}=1$ and $T_2^\textrm{e}/T_1^\textrm{e} = T_2/T_1$, thus { $\eta_\textrm{C}^\textrm{a} = 1-T_2/T_1=\eta_\textrm{C}$; no} effect on the efficiency but the power is reduced by a factor of $\mathcal{A}_K(1+k\tau_1/\gamma_1) \ge 1$. Therefore,
the breaking of the time-scale symmetry is crucial in enhancing the engine performance.
Similar phenomena were found recently in some quantum engines~\cite{Cao,Um}, where the quantum-ness
disappears with the symmetry.

Furthermore, we can see that the active engine can do work with two active reservoirs with the same
effective temperatures, but with different persistence times. This also manifests that the active reservoir should be
characterised not only by its effective temperature but also by its persistent time. More remarkably, the heat flows
can be reversed ($\langle\dot{Q}_1\rangle_\textrm{s} <0$, $\langle\dot{Q}_2\rangle_\textrm{s} >0$ for $T_1^\textrm{e}\ge T_2^\textrm{e}$),
still with the positive work extraction $\langle\dot{W}\rangle_\textrm{s} >0$) when $T_2^\textrm{e}/\mathcal{B}_{K} > T_1^\textrm{e}/\mathcal{A}_{K}$ (see detailed discussions in SM I.C).

To illustrate the enhancement of the active engine performance, we consider a simple case with $\tau_1=0$ (high-temperature equilibrium reservoir) and
$\tau_2>0$ (low-temperature active reservoir). Then, it is clear that $\eta_\textrm{C}^\textrm{a}$ is always larger than
$ \eta_\textrm{C}^\textrm{e}$
as $\mathcal{A}_K =1$ and $\mathcal{B}_K >1$. Furthermore, as $T_1^\textrm{e} = T_1$ and $T_2^\textrm{e}=T_2/(1+ k \tau_2 /\gamma_2) <T_2$, we find $\eta_\textrm{C}^\textrm{a} > \eta_\textrm{C}^\textrm{e} > \eta_\textrm{C}$.
From Eq.~\eqref{eq:x1v2_active}, we can also easily see that the power is enhanced in this
case, compared to the case of both equilibrium reservoirs ($\tau_1 = \tau_2 = 0$).

Figure~\ref{fig:delta-epsilon_plot}(d) shows the region satisfying the stable and useful engine condition with $\tau_1=0$, $\tau_2=0.5$, $T_1^\textrm{e}=2$, and $T_2^\textrm{e}=1$,
which is extended outside of the line of {$\delta=(T_2^\textrm{e}/T_1^\textrm{e})\epsilon$}, where the efficiency $\eta$ is larger than
$\eta_\textrm{C}^\textrm{e}$. The {boundary of the extended region} is given by $\langle x_1 \dot{x}_2 \rangle_\textrm{s} =0$ in Eq.~\eqref{eq:x1v2_active}, thus in this case, {$\delta=(T_2^\textrm{e}/(\mathcal{B}_K T_1^\textrm{e}))\epsilon$}, which is not a straight line because of $K=\delta\epsilon$.
{In Fig.~\ref{fig:delta-epsilon_plot}(e), the normalized efficiency and power as $\tilde{\eta} \equiv \eta/\eta_\textrm{C}^\textrm{e}$ and $\tilde{P} \equiv \langle \dot{W} \rangle_\textrm{s}/P_\textrm{eq}^\textrm{max}$ along the thick dashed line from $\epsilon_1$ to $\epsilon_0$ of Fig.~\ref{fig:delta-epsilon_plot}(d)  are plotted at fixed $\delta=0.8$}, where $\epsilon_0$ is the largest point allowed in the engine region.
The efficiency clearly exceeds the effective Carnot efficiency by far and the power is finite even at
the effective Carnot efficiency. This shows that the conventional power-efficiency tradeoff constraint~\cite{power-eff-rel1,power-eff-rel2,power-eff-rel3}
is not valid in the active engine.

We also show that the EMP of this AOUP engine can surpasses the CA efficiency. Along the curve with fixed $K$, the local maximum power
is obtained at $\delta_\textrm{m}=\sqrt{T_2^\textrm{e}/(\mathcal{B}_K T_1^\textrm{e})} \epsilon_\textrm{m}$ with
the efficiency and the power
\begin{align}
\eta_\textrm{m} = 1 -\sqrt{\frac{T_2^\textrm{e}}{\mathcal{B}_{K} T_1^\textrm{e}}}  > \eta_\textrm{CA}^\textrm{e} ~~~\textrm{and}
~~ \langle \dot{W} \rangle_\textrm{s}^\textrm{m} \equiv \frac{KT_1^\textrm{e} \eta_\textrm{m}^2}{k(\gamma_1+\gamma_2)},
\label{eq:EMP2}
\end{align}
where $\eta_\textrm{CA}^\textrm{e} \equiv 1-\sqrt{T_2^\textrm{e}/T_1^\textrm{e}}$ is the effective CA efficiency. The global power maximum is achieved at a nontrivial value of $K$ for $0\le K \le k^2$ and
$\eta_\textrm{EMP}$
exceeds $\eta_\textrm{CA}^\textrm{e}$.
Figure~\ref{fig:delta-epsilon_plot}(f) shows the plots of $\tilde{\eta}_\textrm{m} \equiv \eta_\textrm{m} /
\eta_\textrm{CA}^\textrm{e}$ and $\tilde{P}_\textrm{m} \equiv \langle \dot{W} \rangle_\textrm{s}^\textrm{m}/P_\textrm{eq}^\textrm{max}$ against $K$. {Note that the dependence of $\eta_C^\textrm{a}$ and $\langle \dot{W} \rangle_\textrm{s}^\textrm{m}$ on general $\tau_1$ and $\tau_2$ is presented in SM III.}

In the above example, it is easy to understand why the efficiency $\eta$ can be larger than $\eta_\textrm{C}$: This is simply because $T_2^\textrm{e}<T_2$ which provides effectively the bigger temperature gradient. However,
there is a nontrivial additional enhancement of the engine performance, which is encoded in the modification factor $\mathcal{M}_{K}$.
In order to understand this remarkable effect, it is useful to resort to a different representation of the equations of motion of the active engine as follows.

It is well known that the AOUP can be mapped on an underdamped Langevin dynamics by introducing an auxiliary velocity $v_i \equiv \dot{x}_i$ and mass $m_i \equiv \gamma_i \tau_i$ as follows~\cite{Fodor,Marconi,Madal}:
\begin{align}
\dot{x}_1 &= v_1,~~m_1 \dot{v}_1 = - k x_1 + \epsilon x_2 + \tau_1 \epsilon v_2 - \Gamma_1  v_1 +\sqrt{2 D_1} \xi_1 ~,  \nonumber \\
\dot{x}_2 &= v_2,~~m_2 \dot{v}_2 = - k x_2 + \delta x_1 + \tau_2 \delta v_1 - \Gamma_2  v_2 +\sqrt{2 D_2} \xi_2 ~, \label{eq:underdamped_Langevin}
\end{align}
which describes the dynamics of a particle in a harmonic trap with the nonconservative force $f_i^\textrm{nc}$ and the unusual
{\em non-antisymmetric} Lorentz-like velocity-dependent force $(\tau_1 \epsilon v_2, \tau_2 \delta v_1)$ in contact with the equilibrium
reservoirs with the temperature ${T}_i^\textrm{e}=D_i/\Gamma_i$.
The standard antisymmetric Lorenz force such as a magnetic force  does not do work by itself. However, the non-antisymmetric Lorentz-like force can do work as well as change the steady-state distribution function in a significant way. Thus, the existence of the velocity-dependent force can promote the work rate as well as the heat rate, which makes it
possible to exceed the effective Carnot efficiency. Note that the heat flow out of the active reservoir in this representation is given
as $\dot{Q}_1^\textrm{a}=v_1\circ(\tau_1\epsilon v_2-\Gamma_1 v_1 +\sqrt{2D_1}\xi_1)$ and similarly for $\dot{Q}_2^\textrm{a}$, of which
the steady-state averages are identical to the heat rates $\langle\dot{Q}_1\rangle_s$ and $\langle\dot{Q}_1\rangle_s$ calculated previously.

It is also useful to study the entropy production (EP) or irreversibility for the active engine.
Some years ago, Zamponi {\em et al.}~showed that the stochastic thermodynamic approach for the EP can be generalized to a non-Markovian process with a memory kernel~\cite{Zamponi}.
Very recently, the EP for the AOUP was explicitly derived using this method, which turns out to be
equivalent to the EP obtained for the above auxiliary underdamped dynamics with the {\em standard} definition
of the parity~\cite{Fodor,Caprini1,Caprini2}. Furthermore, the EP calculation method in an underdamped dynamics with general
velocity-dependent forces is well documented~\cite{Kwon,Lee_old}. In this study, we take this latter approach to derive the EP for the AOUP engine
exactly and show that the {\em unconventional} EP term appearing generally with velocity-dependent forces plays a key role, which provides the main source for the efficiency surpassing the Carnot efficiency in the EP perspective (see SM IV).
\\

\emph{Engine with hybrid reservoirs} --
Finally, we consider a more realistic {\em hybrid} engine by adding active particles (bacteria) into equilibrium fluid
reservoirs~\cite{bacteria_exp4,Bechinger,Dabelow,Caprini2}. Then, the equations of motion are given by
\begin{align}
\gamma_i \dot{x}_i = -\partial_{x_i} \Phi +f_i^\textrm{nc} + \zeta_i + \sqrt{2 \gamma_i T_i} \xi_i^\prime~,
\label{eq:main_eq_hybrid}
\end{align}
where the reservoir noise is composed of two independent noises: a Gaussian white noise $\xi_i^\prime$  with
$\langle \xi_i^\prime (t) \xi_j^\prime (t^\prime) \rangle = \delta_{ij} \delta(t-t^\prime)$
and a Gaussian colored noise $\zeta_i$ with $\langle \zeta_i (t) \zeta_j (t^\prime) \rangle = D_i \delta_{ij} e^{-|t-t^\prime|/\tau_i } /\tau_i$. {Equation~\eqref{eq:main_eq_hybrid} can also describe the dynamics of a self-propelled particle as an engine particle with equilibrium baths~\cite{hybrid}.

In the steady state, the power and heat rates are expressed in the same form as before, e.g.~$\langle \dot{W} \rangle_\textrm{s}=  ( \epsilon - \delta ) \langle x_1 \dot{x}_2 \rangle_\textrm{s} $. Following the previous calculation procedure, we find (see SM I.D)
\begin{align}
\langle x_1 \dot{x}_2 \rangle_\textrm{s} = \frac{  T_1 \delta - T_2 \epsilon }{ k(\gamma_1 +\gamma_2) }+\frac{  1 }{ k(\gamma_1 +\gamma_2) }    \left(  \frac{D_1 \delta }{\Gamma_1 \mathcal{A}_{K}}  - \frac{D_2  \epsilon}{ \Gamma_2 \mathcal{B}_{K}}  \right)~, \label{eq:x1v2_hybrid}
\end{align}
which is a simple sum of two currents due to equilibrium noises and active noises. Note that the power can be enhanced (or reduced) by adding the active noise into the high-temperature (low-temperature) reservoir.
In Figs.~\ref{fig:delta-epsilon_plot}(g), (h), and (i), we plot the engine region, the power, the efficiency with
the parameters $\tau_1=0.5$, $\tau_2=0$, $T_1=2$, $T_2=1$ $D_1=3$, and $D_2=0$.
We also obtain the effective temperature of the hybrid reservoir as ${T}_i^\textrm{e} =T_i+ D_i/\Gamma_i$ (see SM I.B), and then
 the engine condition yields
\begin{align}
0\le \eta \leq 1-\frac{T_2^\textrm{e}+T_2 (\mathcal{B}_K-1)}{T_1^\textrm{e}+T_1(\mathcal{A}_K-1)}\left(\frac{\mathcal{A}_K}{\mathcal{B}_K}\right)\equiv \eta_\textrm{C}^{hybrid} ~,
\label{eq:new_bound_hybrid}
\end{align}
with the maximum efficiency $\eta_\textrm{C}^{hybrid}$ which can again exceed the effective Carnot
efficiency  $\eta_\textrm{C}^\textrm{e}$. The EMP and the maximum power can be also derived.
\\

\emph{Conclusion} -- We demonstrated that the power and the efficiency of a device working in nonequilibrium active environments with Gaussian colored noises with finite persistent time can overcome the conventional Carnot limit. This is possible because the total EP in the steady state cannot be expressed solely by the Clausius EP, 
and the unconventional EP term~\cite{Kwon,Lee_old} emerges due to a velocity-dependent force present in the underdamped representation. In fact, the Clausius EP is negative for overcoming the Carnot bound,
which is compensated by the positive contribution from the unconventional EP. This gives rise to the non-negative total
EP, which is fully consistent with the thermodynamic second law.

We note that our main results should be also applied to more general cases with non-Gaussian colored noises. This implies that the non-Markovianity of the active noise is more crucial than its non-Gaussianity for the out-performance of the active engine, in contrast to the recent claim by Krishnamurthy et al.~\cite{bacterial_bath_exp}. Furthermore, we find that the time-scale symmetry breaking between two active reservoirs is necessary for the supremacy of the active engine.

Our result is readily realizable and applicable to the energy harvesting devices in bacterial or active baths. Thus, our conclusion provides a new way of developing high-performance energy-harvesting devices harnessing energy of microorganisms which exist almost everywhere in nature.

\begin{acknowledgments}
Authors acknowlege the Korea Institute for Advanced Study for providing computing resources (KIAS Center for Advanced Computation Linux Cluster System). This research was supported by the NRF Grant No.~2017R1D1A1B06035497 (HP) and the KIAS individual Grants No.~PG013604 (HP), PG074001 (JMP), QP064902 (JSL) at Korea Institute for Advanced Study.
\end{acknowledgments}

\vfil\eject


\begin{thebibliography}{99}
\bibitem{energy_harvest_review1} R.J.M. Vullers, R. van Schaijk, I. Doms, C. Van Hoof, R. Mertens, Micropower energy harvesting, Solid-State Electronics \textbf{53}, 684–693 (2009).

\bibitem{energy_harvest_review2} L. Mateu and F. Moll, Review of energy harvesting techniques and applications for microelectronics (Keynote Address), Proc. SPIE \textbf{5837}, VLSI Circuits and Systems II, (2005).

\bibitem{graphene} P. M. Thibado, P. Kumar, S. Singh, M. Ruiz-Garcia, A. Lasanta, and L. L. Bonilla, Fluctuation-induced current from freestanding graphene: toward nanoscale energy harvesting, e-print arXiv:2002.09947.

\bibitem{harvesting_thermoelectric} M. Josefsson, A. Svilans, A. M. Burke, E. A. Hoffmann, S. Fahlvik, C. Thelander, M. Leijnse, and H. Linke, A quantum-dot heat engine operating close to the thermodynamic efficiency limits, Nature Nanotech. \textbf{13}, 920 (2018).

\bibitem{harvesting_photo} N. Femia, G. Petrone, G. Spagnuolo, M. Vitelli,
\emph{Power Electronics and Control Techniques for Maximum Energy Harvesting in Photovoltaic Systems}, 1st ed. (CRC Press, 2013).

\bibitem{harvesting_piezo} H. S. Kim, J.-H. Kim, and J. Kim, A review of piezoelectric energy harvesting based on vibration, Int. J. Precis. Eng. Man. \textbf{12}, 1129 (2011).

\bibitem{squeezed1} W. Niedenzu, V. Mukherjee, A. Ghosh, A. G. Kofman, and G. Kurizki, Quantum engine efficiency bound beyond the second law of thermodynamics, Nature Comm.  \textbf{9}, 165 (2018).

\bibitem{squeezed2} J. Klaers, S. Faelt, A. Imamoglu, and E. Togan, Squeezed Thermal Reservoirs as a Resource for a Nanomechanical Engine beyond the Carnot Limit, Phys. Rev. X \textbf{7}, 031044 (2017).

\bibitem{squeezed3} J. Ro{\ss}nagel, O. Abah, F. Schmidt-Kaler, K. Singer, and E. Lutz, Nanoscale Heat Engine Beyond the Carnot Limit, Phys. Rev. Lett. \textbf{112}, 030602 (2014).



\bibitem{bacterial_bath_exp} S. Krishnamurthy, S. Ghosh, D. Chatterji, R. Ganapathy, and A. K. Sood, A micrometre-sized heat engine operating between bacterial reservoirs,
Nature Phys. \textbf{12}, 1134–1138 (2016).
\bibitem{wijland} R. Zakine, A. Solon, T. Gingrich, and F. van Wijland, Stochastic Stirling engine operating in contact with active baths, Entropy \textbf{19}, 193 (2017).



\bibitem{JSLee1} J. S. Lee and H. Park, Carnot efficiency is reachable in an irreversible process, Sci. Rep. \textbf{7}, 10725 (2017).

\bibitem{JSLee2} J. S. Lee, S. H. Lee, J. Um, H. Park, Carnot efficiency and zero-entropy-production rate do not guarantee reversibility of a process, J. Korean Phys. Soc. \textbf{75}, 948 (2019).

\bibitem{PoEs} Polettini and Esposito, Carnot efficiency at divergent power output, EPL \textbf{118}, 40003 (2017).

\bibitem{bacteria_exp1} X.-L. Wu and A. Libchaber, Particle diffusion in a quasi-two-dimensional bacterial bath, Phys. Rev. Lett. \textbf{84}, 3017 (2000).

\bibitem{bacteria_exp2} K. C. Leptos, J. S. Guasto, J. P. Gollub, A. I. Pesci, and R. E. Goldstein, Dynamics of enhanced tracer diffusion in suspensions of swimming Eukaryotic microorganisms, Phys. Rev. Lett. \textbf{103}, 198103 (2009).

\bibitem{bacteria_exp3} H. Kurtuldu, J. S. Guasto, K. A. Johnson, and J. P. Gollub, Enhancement of biomixing by swimming algal cells in two-dimensional films, PNAS \textbf{108}, 10391-10395 (2011).

\bibitem{bacteria_exp4} C. Maggi, M. Paoluzzi, N. Pellicciotta, A. Lepore, L. Angelani, and R. Di Leonardo, Generalized energy equipartition in harmonic oscillators driven by active baths, Phys. Rev. Lett. \textbf{113}, 238303 (2014).
\bibitem{Bechinger} C. Bechinger, R. Di Leonardo, H. L\"owen, C. Reichhardt, G. Volpe, and G. Volpe, Active particles in complex and crowded environments, Rev. Mod. Phys. \textbf{88}, 045006 (2016).
\bibitem{Dabelow} L. Dabelow, S. Bo, and R. Eichhorn, Irreversibility in Active Matter Systems: Fluctuation Theorem and Mutual Information, Phys. Rev. X \textbf{9}, 021009 (2019).



\bibitem{power-eff-rel1} Naoto Shiraishi, Keiji Saito, Hal Tasaki, Universal Trade-Off Relation between Power and Efficiency for Heat Engines, Phys. Rev. Lett. \textbf{117}, 190601 (2016).

\bibitem{power-eff-rel2} A. Dechant and S.-I. Sasa, Entropic bounds on currents in Langevin systems, Phys. Rev. E \textbf{97}, 062101 (2018).

\bibitem{power-eff-rel3} P. Pietzonka and U. Seifert, Universal Trade-Off between Power, Efficiency, and Constancy in Steady-State Heat Engines, Phys. Rev. Lett. \textbf{120}, 190602 (2018).

\bibitem{CAefficiency} F. L. Curzon, and B. Ahlborn, Efficiency of a Carnot engine at maximum power output, Am. J. Phys. \textbf{43}, 22 (1975).


\bibitem{Crisanti} A. Crisanti, A. Puglisi, and D. Villamaina, Nonequilibrium and information: The role of cross correlations, Phys.
Rev. E \textbf{85}, 061127 (2012).

\bibitem{ParkJM} J.-M. Park, H.-M. Chun, and J. D. Noh, Efficiency at maximum power and efficiency fluctuations in a linear Brownian heat-engine model, Phys. Rev. E \textbf{94}, 012127 (2016).



\bibitem{Pietzonka} P. Pietzonka and U. Seifert, Universal Trade-Off between Power, Efficiency, and Constancy in Steady-State Heat Engines, Phys. Rev. Lett. \textbf{120}, 190602 (2018).

\bibitem{Chulan} C. Kwon,  J. D. Noh, and H. Park, Nonequilibrium fluctuations for linear diffusion dynamics, Phys. Rev. E \textbf{83}, 061145 (2011).

\bibitem{Chun} H.-M. Chun, L. P. Fischer, and U. Seifert, Effect of a magnetic field on the thermodynamic uncertainty relation, Phys. Rev. E \textbf{99}, 042128 (2019).



\bibitem{Filliger} R. Filliger and P. Reimann, Brownian Gyrator: A Minimal Heat Engine on the Nanoscale, Phys. Rev. Lett. \textbf{99}, 230602 (2007).

\bibitem{Chiang} K.-H. Chiang, C.-L. Lee, P.-Y. Lai, and Y.-F. Chen, Electrical autonomous Brownian gyrator, Phys. Rev. E \textbf{96}, 032123 (2017).

\bibitem{Risken} H. Risken, \emph{The Fokker-Planck Equation, Methods of Solution and Applications} (Springer, 1996).

\bibitem{Cao} J. Thingna, D. Manzano, and J. Cao, Dynamical signatures of molecular symmetries in nonequilibrium quantum transport,
Sci. Rep. \textbf{6}, 28027 (2016).
\bibitem{Um} J. Um, K. Dorfman, and H. Park, Coherence effect in a multi-level quantum-dot heat engine (unpublished).


\bibitem{Fodor} \'{E}. Fodor, C. Nardini, M. E. Cates, J. Tailleur, P. Visco, and F. van Wijland, How far from equilibrium is active matter?, Phys. Rev. Lett. \textbf{117}, 038103 (2016).

\bibitem{Marconi} U. M. B. Marconi, A. Puglisi, and C. Maggi, Heat, temperature and Clausius inequality in a model for active Brownian particles, Sci. Rep. \textbf{7}, 46496 (2017).

\bibitem{Madal} D. Mandal, K. Klymko, and M. R. DeWeese, Entropy Production and Fluctuation Theorems for Active Matter, Phys. Rev. Lett. \textbf{119}, 258001 (2017).

\bibitem{Zamponi} F. Zamponi, F. Bonetto, L. F. Cugliandolo, and J. Kurchan, A fluctuation theorem for
non-equilibrium relaxational systems driven by external forces, J. Stat. Mech. P09013 (2005).

\bibitem{Caprini1} L. Caprini, U. M. B. Marconi, A. Puglisi, and A. Vulpiani,
Comment on “Entropy production and fluctuation theorems for active matter", Phys. Rev. Lett. \textbf{121}, 139801 (2018).

\bibitem{Caprini2} L. Caprini, U. M. B. Marconi, A. Puglisi, and A. Vulpiani,
The entropy production of Ornstein-Uhlenbeck active particles: a path integral method for correlations,
J. Stat. Mech. 053203 (2019).


\bibitem{Kwon} C. Kwon, J. Yeo, H. K. Lee, and H. Park, Unconventional entropy production in the presence of
momentum-dependent forces, J. Korean Phys. Soc. \textbf{68}, 633 (2016).

\bibitem{Lee_old} H. K. Lee, S. Lahiri, and H. Park, Nonequilibrium steady states in Langevin thermal systems, Phys. Rev. E
\textbf{96}, 022134 (2017).

\bibitem{hybrid} T. F. F. Farage, P. Krinninger, and J. M. Brader, Effective interactions in active Brownian suspensions, Phys. Rev E \textbf{91}, 042310 (2015).







\end{thebibliography}
\end{document}


\title{Supplemental Material for ``Brownian heat engine with active reservoirs''}
\author{Jae Sung Lee}
\author{Jong-Min Park}
\author{Hyunggyu~Park}
\affiliation{School of Physics and Quantum Universe Center, Korea Institute for Advanced Study, Seoul 02455, Korea}

\newcommand{\red}[1]{{\color{red}#1}}


\maketitle

\section{Covariant matrix, Steady-state distribution, Work and Heat rates in the Ornstein-Uhlenbeck process}

 The multivariate Ornstein-Uhlenbeck process~\cite{Risken,Chulan} is described by
\begin{align}
d\textbf{z} = - \textsf{A} \textbf{z} dt + d\Xi \label{eqA:OU},
\end{align}
where $\textbf{z}$ is a $d$-dimensional state (column) vector and $\mathsf{A}$ is a linear force matrix. The noise vector $d \Xi$ satisfies
$\langle d\Xi (t)\rangle =0$ and $\langle d\Xi(t) d\Xi^{\textsf{T}} (t)   \rangle = 2\textsf{D} dt$, where $\textsf{D}$ is a symmetric diffusion matrix. Note that the superscript `$\textsf{T}$' represents the transpose.

The covariant matrix $\Sigma$  is defined as
\begin{equation}
\Sigma = \langle \textbf{z} \textbf{z}^{\textsf{T}} \rangle_\textrm{s} =\Sigma^{\textsf{T}}~,
\end{equation}
where $\langle\cdots\rangle_\textrm{s}$ denotes the steady-state average. Its incremental satisfies
\begin{align}
d \Sigma = \langle d\textbf{z}\circ \textbf{z}^\textsf{T}\rangle_\textrm{s} +\langle \textbf{z}\circ d\textbf{z}^\textsf{T}\rangle_\textrm{s}
= [-\textsf{A} \Sigma - \Sigma \textsf{A}^{\textsf{T}} + 2\textsf{D}]dt  =0~,
\end{align}
with the Stratonovich multiplication $\circ$, thus we obtain
\begin{equation}
\textsf{A} \Sigma + \Sigma \textsf{A}^{\textsf{T}} = 2\textsf{D}~. \label{eqS:OUsln}
\end{equation}
Similarly, we find
\begin{align}
\langle \textbf{z}\circ \dot{\textbf{z}}^\textsf{T}\rangle_\textrm{s}
=-\Sigma\textsf{A}^{\textsf{T}}+\textsf{D}~, \label{eqS:OUrate}
\end{align}
which is useful for calculation of work and heat rates.

The steady-state distribution is given as~\cite{Risken,Chulan}
\begin{align}
P(\textbf{z})=\frac{1}{\sqrt{|2\pi\Sigma|}} \exp\left[{-\frac{1}{2}\mathbf{z}^{\mathsf{T}}\Sigma^{-1} \mathbf{z}}\right]~, \label{eqS:OUss}
\end{align}
where $|\cdots|$ is the determinant. The stability condition for the steady state is obtained by requiring the positivity of all eigenvalues of the covaraint matrix $\Sigma$, or equivalently of the linear force matrix $\mathsf{A}$ for the non-negative diagonal diffusion matrix $\mathsf{D}$.

\subsection{equilibrium-reservoir engine}

The equations of motion are given in Eqs.~(1) and (2) of the main text.
Then, the state vector is defined as $\textbf{z} =(x_1, x_2)^{\textsf{T}}$ with
\begin{align}
\textsf{A}= \left( \begin{array}{cc}
\frac{k}{\gamma_1} & -\frac{\epsilon}{\gamma_1}  \\
-\frac{\delta}{\gamma_2} & \frac{k}{\gamma_2}  \\
\end{array} \right)~,~~
\textsf{D}= \left( \begin{array}{cc}
\frac{T_1}{\gamma_1} & 0  \\
0 & \frac{T_2}{\gamma_2}  \\
\end{array} \right)~.
\end{align}

Solving Eq.~\eqref{eqS:OUsln}, we find
\begin{align}
\Sigma_{11} &= \langle x_1^2 \rangle_\textrm{s} = \frac{ T_1 [k^2\gamma_1 + (k^2-K)\gamma_2]+ T_2\epsilon^2\gamma_2 }{ k(\gamma_1 +\gamma_2)(k^2 -K) }, \nonumber \\
\Sigma_{12} &=\langle x_1 x_2 \rangle_\textrm{s} = \frac{ T_1\delta \gamma_1 + T_2 \epsilon\gamma_2  }{ (\gamma_1 +\gamma_2)(k^2 -K) }, \nonumber \\
\Sigma_{22} &=\langle x_2^2 \rangle_\textrm{s} = \frac{
 T_1 \delta^2\gamma_1 +T_2 [k^2\gamma_2 + (k^2-K)\gamma_1] }{ k(\gamma_1 +\gamma_2)(k^2 -K) }~, \label{eqS:slns}
\end{align}
with $K=\delta\epsilon$. Note that $\Sigma$ is invariant under the time rescaling as $\gamma_i\rightarrow a\gamma_i$ with fixed $T_i$.
With these results, we get from Eq.\eqref{eqS:OUrate}
\begin{align}
\langle x_1 \dot{x}_2 \rangle_\textrm{s} = \left(-\Sigma\textsf{A}^\textsf{T}+\textsf{D}\right)_{12}=-\Sigma_{11}\textsf{A}_{21}-
\Sigma_{12}\textsf{A}_{22}= \frac{ T_1 \delta - T_2 \epsilon }{ k(\gamma_1 +\gamma_2) }. \label{x1v2}
\end{align}

For stability, we diagonalize $\textsf{A}$ with two eigenvalues as
\begin{align}
\lambda^{\pm} = \frac{ k(\gamma_1 +\gamma_2) \pm \sqrt{ k^2 (\gamma_1+\gamma_2)^2 +4\gamma_1 \gamma_2 ( K - k^2) } }{2 \gamma_1 \gamma_2}. \label{eqS:eigenvalues}
\end{align}
For a stable solution, the real part of all eigenvalues of $\textsf{A}$ should be positive. Thus, the stability condition is
given by
\begin{align}
 K  < k^2. \label{eqS:stability_cond}
\end{align}

\subsection{effective temperature of a single active or hybrid reservoir}

We consider an one-dimensional overdamped linear system in contact with a single hybrid reservoir as
\begin{align}
\gamma \dot{x} =-kx +\zeta+\sqrt{2\gamma T}\xi^\prime, \quad \tau\dot{\zeta}=-\zeta +\sqrt{2D}\xi~,
\end{align}
with $\langle\xi^\prime(t)\xi^\prime(t')\rangle=\delta(t-t')$ and $\langle\xi(t)\xi(t')\rangle=\delta(t-t')$.
One can recover the active-reservoir case by setting $T=0$.
With the state vector $\textbf{z} =(x, \zeta)^{\textsf{T}}$, we have
\begin{align}
\textsf{A}= \left( \begin{array}{cc}
\frac{k}{\gamma} & -\frac{1}{\gamma}  \\
0 & \frac{1}{\tau}  \\
\end{array} \right)~,~~
\textsf{D}= \left( \begin{array}{cc}
\frac{T}{\gamma} & 0  \\
0 & \frac{D}{\tau^2}  \\
\end{array} \right)~.
\end{align}

Then, Eq.~\eqref{eqS:OUsln} yields
\begin{align}
\Sigma_{11} = \langle x^2 \rangle_\textrm{s} = \frac{1}{k}\left(T+\frac{D}{\Gamma}\right),~~\Sigma_{12} =\langle x \zeta \rangle_\textrm{s} = \frac{D}{\Gamma},~~\Sigma_{22} =\langle \zeta^2\rangle_\textrm{s} = \frac{ D}{\tau}~~~\textrm{with}~~\Gamma=\gamma+k\tau~.\label{eqS:actives}
\end{align}
Inverting $\Sigma$, we get the steady state given by Eq.~\eqref{eqS:OUss} in terms of $\textbf{z}$.
The steady state of the true dynamic variable $x$ can be obtained by integrating Eq.~\eqref{eqS:OUss} over $\zeta$ as
\begin{align}
P(x)=\frac{1}{\sqrt{2\pi \Sigma_{11}}} \exp\left[-\frac{x^2}{2\Sigma_{11}}\right]=\sqrt{\frac{k}{2\pi T^\textrm{e}}} \exp\left[-\frac{kx^2}{2T^\textrm{e}}\right]~, \label{eqS:activess}
\end{align}
with the effective temperature  $T^\textrm{e}=T+D/\Gamma$. For the active engine without
a nonconservative force (no coupling between two engine degrees of freedom, $x_1$ and $x_2$),
 we get Eq.~(11)  for the stationary distribution with $T=0$.

\subsection{active-reservoir engine}

The equations of motion for the active-reservoir engine are given in Eqs.~(9) and (10).
The state vector is defined as $\textbf{z} =(x_1, x_2,
\zeta_1, \zeta_2)^{\textsf{T}}$ with
\begin{align}
\textsf{A}= \left( \begin{array}{cccc}
\frac{k}{\gamma_1} & -\frac{\epsilon}{\gamma_1} & -\frac{1}{\gamma_1} & 0 \\
-\frac{\delta}{\gamma_2} & \frac{k}{\gamma_2}  & 0 & -\frac{1}{\gamma_2} \\
0 & 0 & \frac{1}{\tau_1} & 0 \\
0 & 0 & 0 & \frac{1}{\tau_2} \\
\end{array} \right)~,~~
\textsf{D}= \left( \begin{array}{cccc}
0 & 0 & 0 & 0 \\
0 & 0 & 0 & 0 \\
0 & 0 & \frac{D_1 }{\tau_1^2} & 0 \\
0 & 0 & 0 & \frac{D_2 }{\tau_2^2} \\
\end{array} \right).\label{AD_active}
\end{align}

Solving Eq.~\eqref{eqS:OUsln}, we find
\begin{align}
\Sigma_{11} &= \frac{1}{k(\gamma_1 + \gamma_2)(k^2 - K)}
 \left[ \frac{D_1\{k^2(\gamma_2\Gamma_1+k\gamma_1\tau_1)+(k^2-K)\gamma_2^2\}}{\gamma_2 \Gamma_1 \mathcal{A}_{K}}
 + \frac{\epsilon^2 D_2(\gamma_1\Gamma_2+k\gamma_2\tau_2)}{\gamma_1 \Gamma_2 \mathcal{B}_{K}} \right]~,  \nonumber \\
\Sigma_{12} &= \frac{1}{(\gamma_1 + \gamma_2)(k^2 - K)}
 \left[ \frac{\delta D_1 (\gamma_2\Gamma_1+k\gamma_1\tau_1)}{\gamma_2 \Gamma_1 \mathcal{A}_{K}}
 + \frac{\epsilon D_2(\gamma_1\Gamma_2+k\gamma_2\tau_2)}{\gamma_1 \Gamma_2 \mathcal{B}_{K}} \right]~,  \nonumber \\
\Sigma_{22} &= \frac{1}{k(\gamma_1 + \gamma_2)(k^2 - K)}
 \left[ \frac{\delta^2 D_1(\gamma_2\Gamma_1+k\gamma_1\tau_1)}{\gamma_2 \Gamma_1 \mathcal{A}_{K}}
 +  \frac{D_2\{k^2(\gamma_1\Gamma_2+k\gamma_2\tau_2)+(k^2-K)\gamma_1^2\}}{\gamma_1 \Gamma_2 \mathcal{B}_{K}} \right]~,  \nonumber \\
\Sigma_{13} &= \frac{D_1(\gamma_2 +k\tau_1)}{\gamma_2 \Gamma_1 \mathcal{A}_{K}}, \quad
\Sigma_{14} = \frac{\epsilon D_2\tau_2}{\gamma_1 \Gamma_2 \mathcal{B}_{K}},  \quad
\Sigma_{23} = \frac{\delta D_1\tau_1}{\gamma_2 \Gamma_1 \mathcal{A}_{K}}, \quad
\Sigma_{24} = \frac{D_2(\gamma_1+k\tau_2)}{\gamma_1 \Gamma_2 \mathcal{B}_{K}}, \nonumber \\
\Sigma_{33} &= \frac{D_1}{\tau_1}, \quad
\Sigma_{34} = 0,  \quad
\Sigma_{44} = \frac{D_2}{\tau_2}, \label{CM_active}
\end{align}
where $\mathcal{A}_{K}$ and $\mathcal{B}_{K}$ are given in Eq.~(13)  with $K=\delta\epsilon$ and
$\Gamma_i=\gamma_i+k\tau_i$. As expected, $\Sigma$ is invariant under the time rescaling as
$\gamma_i\rightarrow a\gamma_i$, $D_i\rightarrow aD_i$, and $\tau_i\rightarrow a\tau_i$, .

From Eq.\eqref{eqS:OUrate}, we get
\begin{align}
\langle x_1 \dot{x}_2 \rangle_\textrm{s} &= \left(-\Sigma\textsf{A}^\textsf{T}+\textsf{D}\right)_{12}=
-\Sigma_{11}\textsf{A}_{21}-
\Sigma_{12}\textsf{A}_{22}-\Sigma_{14}\textsf{A}_{24}~,\nonumber\\
&= \frac{ 1 }{ k(\gamma_1 +\gamma_2) }   \left[  \frac{D_1 \delta }{\Gamma_1 \mathcal{A}_{K}}  - \frac{D_2  \epsilon}{ \Gamma_2 \mathcal{B}_{K}}  \right]= \frac{ 1 }{ k(\gamma_1 +\gamma_2) }   \left[  \frac{T_1^\textrm{e} \delta }{ \mathcal{A}_{K}}  - \frac{T_2^\textrm{e}  \epsilon}{ \mathcal{B}_{K}}  \right]~, \label{eqS:x1v2_active}
\end{align}
with the effective temperatures $T^\textrm{e}_i=D_i/\Gamma_i$.
The stability condition is the same as before in Eq.~\eqref{eqS:stability_cond}, which can be clearly seen from the force
matrix $\mathsf{A}$ in Eq.~\eqref{AD_active}.

There are some interesting cases considered. First, in the case of the equilibrium reservoirs $(\tau_1=\tau_2=0)$, we have $\mathcal{A}_{K}=\mathcal{B}_{K}=1$ and $T_i^\textrm{e}=T_i=D_i/\gamma_i$, then the conventional result, Eq.~(5), is recovered. For $T_1=T_2$, $\langle\dot{W}\rangle_\textrm{s}=(\epsilon-\delta)\langle x_1\dot{x}_2\rangle_\textrm{s}=-(\delta-\epsilon)^2T_1/[k(\gamma_1+\gamma_2)]\le 0$, which implies that
no positive work extraction is  possible with the same-temperature equilibrium reservoirs. Note that the time scale does not enter here, i.e., we get the same result with $\gamma_1=a\gamma_2$ and $D_1=aD_2$ for general positive $a$.

Second, we consider the case with the same effective temperatures ($T_1^\textrm{e}=T_2^\textrm{e}$) with different time scales such that
$\tau_1=a\tau_2$, $\gamma_1=a\gamma_2$, and $D_1=a D_2$. As $\mathcal{A}_{K}\neq \mathcal{B}_{K}$ except for $a=1$, the positive work extraction is possible ($\langle\dot{W}\rangle_\textrm{s} >0$) for $a<1$ where $\mathcal{B}_{K}> \mathcal{A}_{K}$.
This implies that the active reservoir with a shorter time scale is effectively hotter than the active reservoir with a longer one.
Thus, in order to understand the thermodynamics, the
active reservoir must be characterized not only by the effective temperature but also by the time scale.
At $a=1$, two active reservoirs become identical (or considered as a single active reservoir) and the work production is non-positive;
$\langle\dot{W}\rangle_\textrm{s}\sim -(\delta-\epsilon)^2 \le 0$. The $a>1$ case will be discussed later.

Third, we consider the case with the same time scale ($\gamma_1=\gamma_2$ and $\tau_1=\tau_2$) with different
noise strengths ($D_1> D_2$). So, the effective temperatures $T_1^\textrm{e}>T_2^\textrm{e}$. The {\em time scale symmetry }between two active
reservoirs implies not only the same persistence time but also the same physical property of the reservoir-system contact for two active reservoirs. In this case, $\Gamma_1=\Gamma_2$,
$\mathcal{A}_{K}= \mathcal{B}_{K}$,  and  all energetic quantities are essentially the same as those in the equilibrium-reservoir
case with the effective temperatures except for the overall factor $1/\mathcal{A}_{K}$. Thus, the efficiency is bounded by the
(effective) Carnot efficiency; $\eta\le \eta_\textrm{C}^\textrm{e}=1-T_2^\textrm{e}/T_1^\textrm{e}=1-T_2/T_1=\eta_\textrm{C}$. In order to surpass the Carnot efficiency, this symmetry should be broken. In a very recent paper~\cite{wijland} studying a stochastic Stirling engine with active baths, only the time-scale symmetric case was considered and no better efficiency was found with these simple active reservoirs. If the symmetry is broken in this Stirling engine model, we can show that the efficiency for the active reservoirs can surpass that for the equilibrium reservoirs~\cite{JSL10}. The importance of the symmetry
was also pointed out recently in some quantum heat engines in order to enhance the engine performance~\cite{Cao,Um}.

Finally, it is remarkable to see that the heat flow can be reversed when $T_2^\textrm{e}/\mathcal{B}_{K} > T_1^\textrm{e}/\mathcal{A}_{K}$ with $T_1^\textrm{e}\ge T_2^\textrm{e}$ (including the $a>1$ case in the above) and the work extraction is still positive. In the equilibrium-reservoir case, one can devise a refrigerator instead of a heat engine ($\langle \dot{Q}_1\rangle_\textrm{s} <0$ and $\langle \dot{Q}_2\rangle_\textrm{s} >0$), but
the thermodynamic second law with the Clausius entropy production demands the work input ($\langle\dot{W}\rangle_\textrm{s} <0$).
Thus, one should do work on the system in order to extract heat out of the low-temperature reservoir. However, in the active-reservoir engine, it is possible to make the heat flow out of the low effective-temperature reservoir ($\langle \dot{Q}_2\rangle_\textrm{s} >0$) and do work outside ($\langle\dot{W}\rangle_\textrm{s} >0$) and the remaining heat dissipates into the high effective-temperature reservoir ($\langle\dot{Q}_1\rangle_\textrm{s} <0$.) See Fig.~S1.
In this case, one should define the efficiency as
$\eta=\langle \dot{W}\rangle_\textrm{s}/\langle \dot{Q}_2\rangle_\textrm{s}=1-\epsilon/\delta$. We note that
this is fully consistent with the thermodynamic second law due to the presence of an {\em unconventional} EP term~\cite{Kwon,Lee_old}, which will be discussed later in SM IV.

\begin{figure}
\centering
\includegraphics[width=0.5\linewidth]{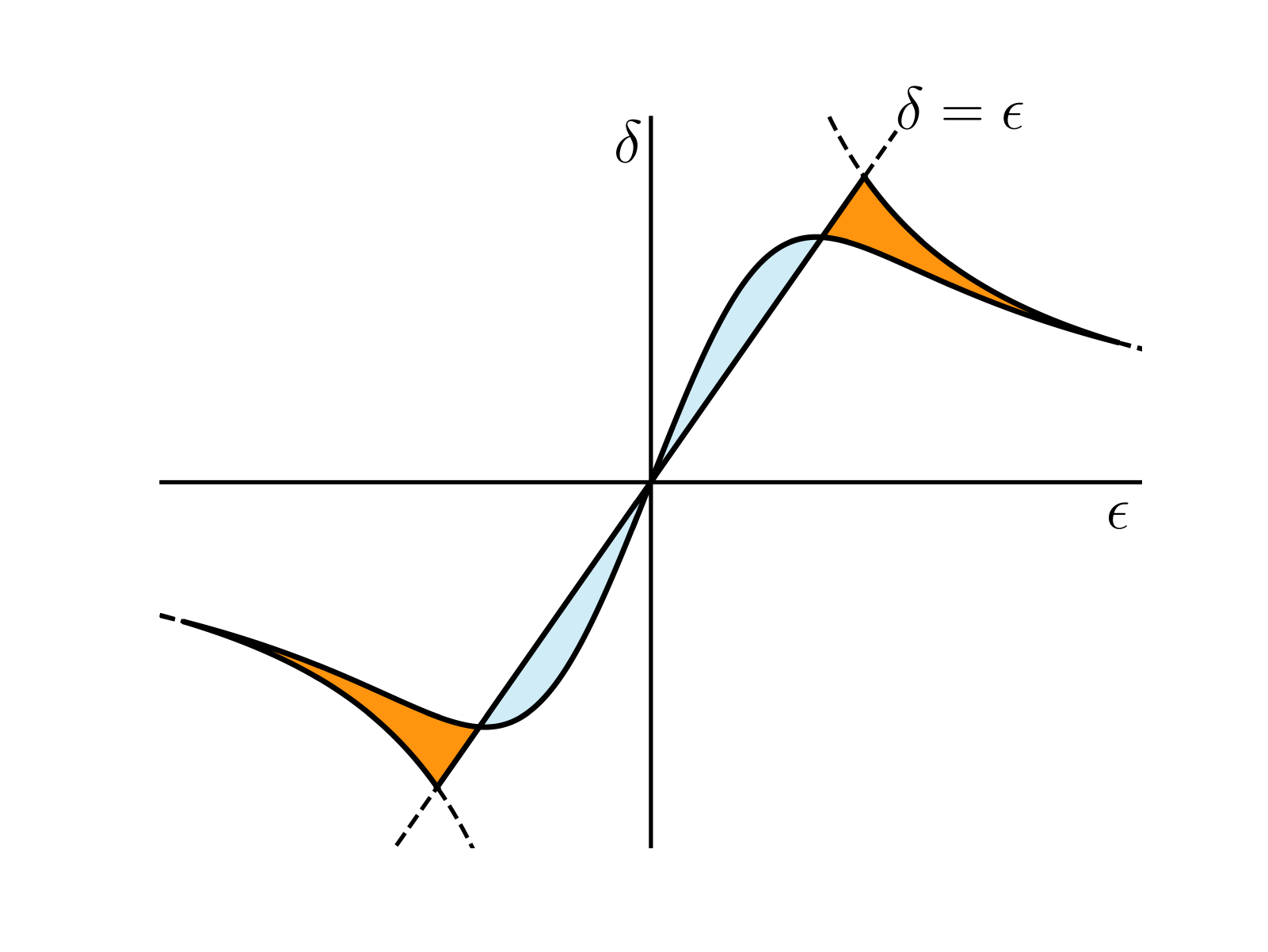}
\caption{Engine with active reservoirs with the reverse heat flows.
Shaded areas, enclosed by the curves
$\delta = \epsilon$,
$\epsilon \delta = k^2$, and
$T_1^\textrm{e} \delta / \mathcal{A}_K =
T_2^\textrm{e} \epsilon / \mathcal{B}_K$, denote the areas in which the device works as
a stable and useful engine with positive work extraction. The orange-colored region denotes the normal-engine area, while the sky-blue-colored region
denotes the engine area with the reversed heat flows.
For this plot,
$\tau_1 = 16$, $\gamma_1 = 0.1$, $D_1 = 7$, and
$k = \tau_2 = \gamma_2 = D_2 = 1$ are used.
} \label{fig:Lij}
\end{figure}

\subsection{hybrid-reservoir engine}

The equations of motion for the hybrid-reservoir engine are given in Eqs.~(17) and (2).
The state vector is defined as $\textbf{z} =(x_1, x_2,
\zeta_1, \zeta_2)^{\textsf{T}}$ with
\begin{align}
\textsf{A}= \left( \begin{array}{cccc}
\frac{k}{\gamma_1} & -\frac{\epsilon}{\gamma_1} & -\frac{1}{\gamma_1} & 0 \\
-\frac{\delta}{\gamma_2} & \frac{k}{\gamma_2}  & 0 & -\frac{1}{\gamma_2} \\
0 & 0 & \frac{1}{\tau_1} & 0 \\
0 & 0 & 0 & \frac{1}{\tau_2} \\
\end{array} \right)~,~~
\textsf{D}= \left( \begin{array}{cccc}
\frac{T_1}{\gamma_1} & 0 & 0 & 0 \\
0 & \frac{T_2}{\gamma_2} & 0 & 0 \\
0 & 0 & \frac{D_1 }{\tau_1^2} & 0 \\
0 & 0 & 0 & \frac{D_2 }{\tau_2^2} \\
\end{array} \right).\label{AD_hybrid}
\end{align}

Solving Eq.~\eqref{eqS:OUsln}, we find that the covariant matrix is given by the simple sum of the covariant matrices for the equilibrium reservoirs in Eq.~\eqref{eqS:slns} and for the active reservoirs in Eq.~\eqref{CM_active}. The matrix elements for $\Sigma_{13}$, $\Sigma_{14}$, $\Sigma_{23}$, $\Sigma_{24}$, $\Sigma_{33}$, $\Sigma_{34}$, and $\Sigma_{44}$ are identical to those for the active reservoirs in Eq.~\eqref{CM_active}.
From Eq.\eqref{eqS:OUrate}, we easily get
\begin{align}
\langle x_1 \dot{x}_2 \rangle_\textrm{s} &= \left(-\Sigma\textsf{A}^\textsf{T}+\textsf{D}\right)_{12}=
-\Sigma_{11}\textsf{A}_{21}-
\Sigma_{12}\textsf{A}_{22}-\Sigma_{14}\textsf{A}_{24}~,\nonumber\\
&= \frac{ 1 }{ k(\gamma_1 +\gamma_2) }   \left[ T_1 \delta - T_2 \epsilon +  \frac{D_1 \delta }{\Gamma_1 \mathcal{A}_{K}}  - \frac{D_2  \epsilon}{ \Gamma_2 \mathcal{B}_{K}}  \right]~, \label{eqS:x1v2_hybrid}
\end{align}
which is also the simple sum of Eqs.~\eqref{x1v2} and \eqref{eqS:x1v2_active}.
The stability condition is the same as before in Eq.~\eqref{eqS:stability_cond}, which can be clearly seen from the force
matrix $\mathsf{A}$ in Eq.~\eqref{AD_hybrid}.

\section{definition of the efficiency for an active or hybrid engine}

In this study, we define the efficiency of an active (hybrid) engine as the conventional one for a passive engine, i.e., a fraction of the extracted work out of the heat energy transferred from the hot reservoir.
In order to keep the steadiness of the hot reservoir, the same amount of energy should be supplied to the reservoir from outside, which can be consider as a {\em cost} from the energy perspective.
For an  active engine, there should be an additional cost to maintain the activeness (persistent time) of the reservoir.  This can be realized by feeding the bacteria with proper food. It is, however, difficult to
estimate this extra cost in a systematic way from the energy perspective, because the chemical energy conversion into the bacteria activity (persistent time) is too complicated and species-dependent.
Nevertheless, the feeding is expected to be not costly (almost free) as we can witness in our daily life that many bacteria can proliferate in the water by themselves without a `practical cost' because microorganisms and their food are everywhere and abundant in nature, especially in biological environments at the cellular level. Thus, this additional cost may be negligible in a practical sense and our definition of the efficiency is still
meaningful and regarded as a 'practical efficiency', i.e., the fraction of the extracted work out of the practical cost. In a more strict sense, our efficiency can be considered as an upper bound of the practical efficiency. In contrast, the quantum engines with squeezed reservoirs need an extra energy expended by laser to maintain the squeezed state, where this additional cost to maintain nonequilibrium-ness can be readily addressed.

\section{power and efficiency for an active-reservoir engine}

The activeness of a reservoir $i$ is characterized by the persistent time $\tau_i$. A non-zero $\tau_i$ lowers the reservoir temperature
effectively as $T_i^e=T_i/[1+k\tau_i/\gamma_i]<T_i$ by itself and increases the value of the factor $\mathcal{A}_K$ ($\mathcal{B}_K$) when it is combined with another reservoir (see Eq.~(13) of the main text). Note that both of $\mathcal{A}_K$ and $\mathcal{B}_K$ monotonically increase with $\tau_1$ and $\tau_2$, respectively.

\begin{figure}
\centering
\includegraphics[width=0.329\linewidth]{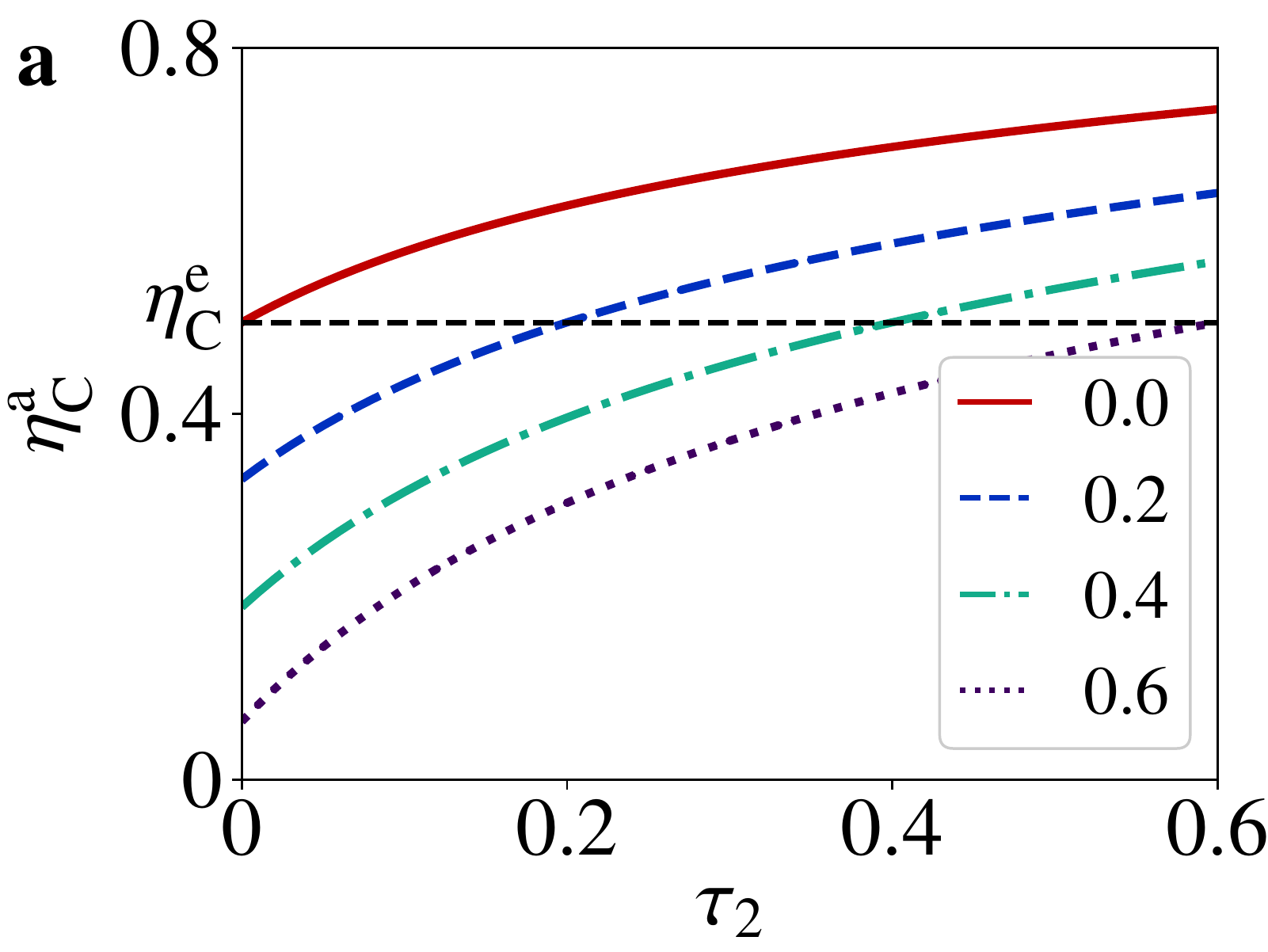}
\includegraphics[width=0.329\linewidth]{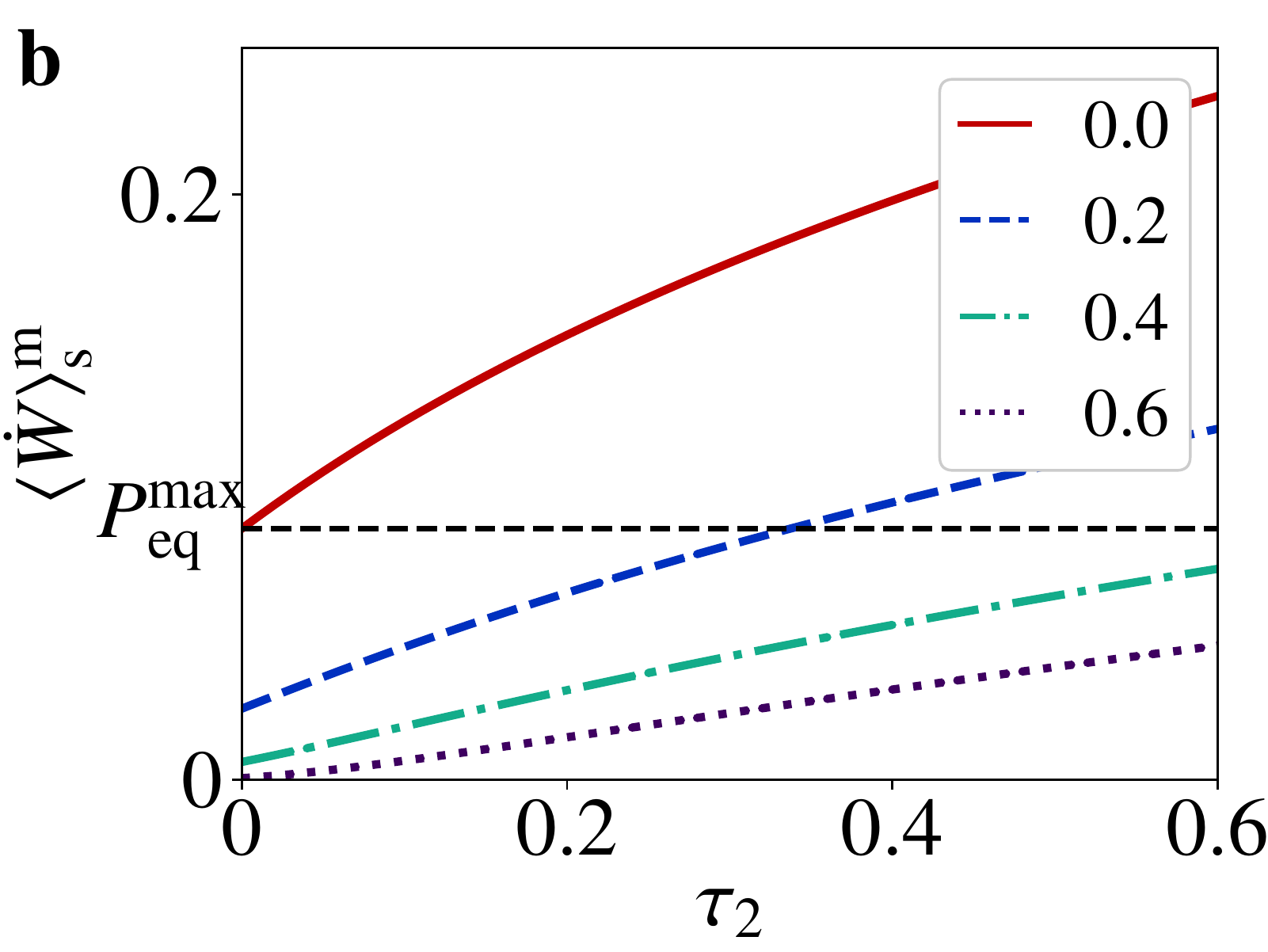}
\includegraphics[width=0.329\linewidth]{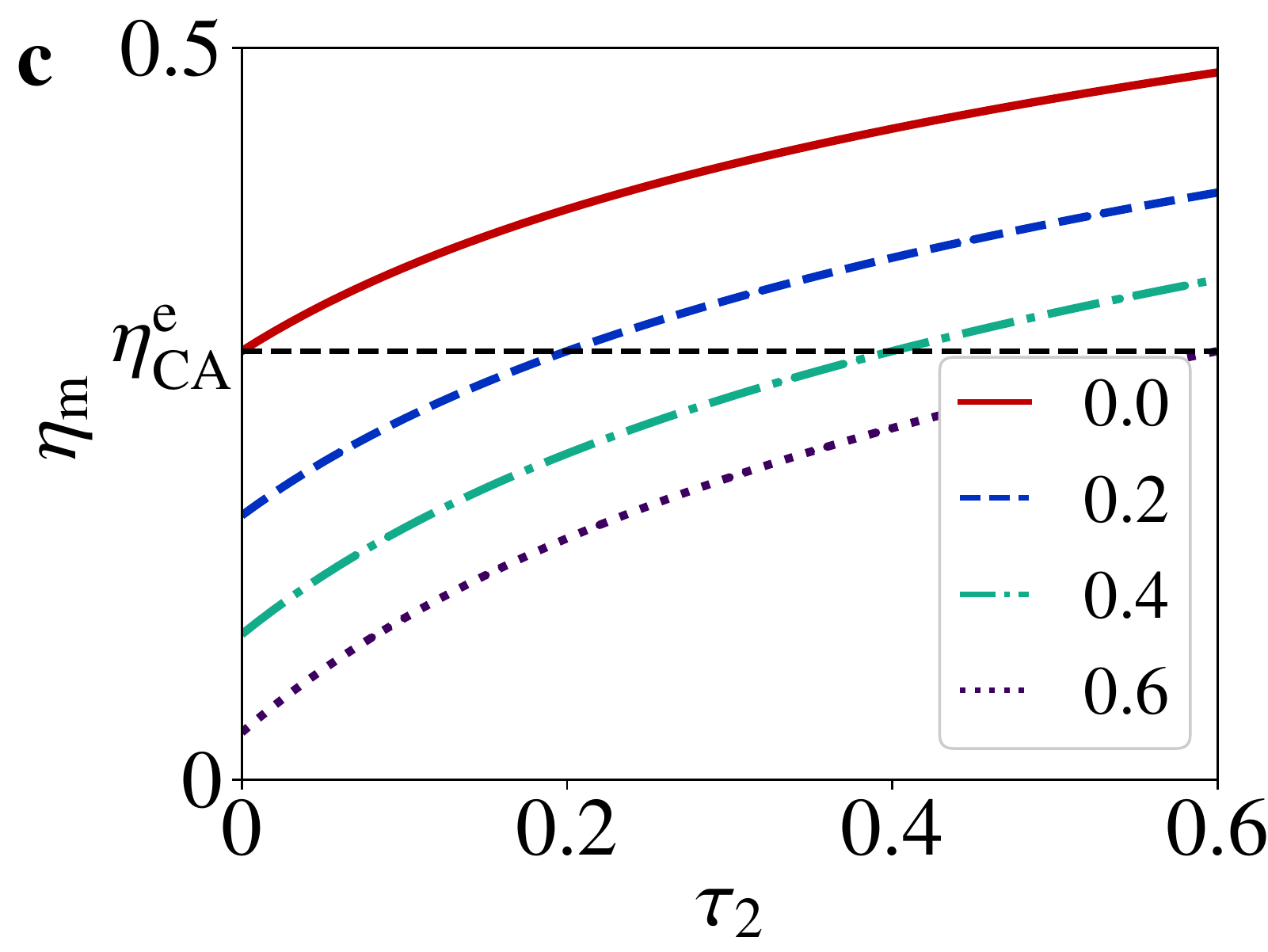}
\caption{
$\tau_2$-dependent behavior of (a) the maximum efficiency
$\eta_\textrm{C}^\textrm{a}$,
(b) the local maximum power
$\langle \dot{W} \rangle_\textrm{s}^\textrm{m}$, and
(c) the efficiency at the maximum power
$\eta_\textrm{m}$, for
$\tau_1 = 0$ (red solid curve),
$0.2$ (blue dashed curve),
$0.4$ (cyan dash-dotted curve),
and
$0.6$ (magenta dotted curve).
Black dashed lines indicate the conventional limit of a passive engine with
the effective temperatures.
} \label{fig:etaC_power_emp}
\end{figure}

The power and efficiency of an active engine depend on the activeness of both hot and cold reservoirs. The maximum efficiency
$\eta_\textrm{C}^\textrm{a}$  is also a function of $\tau_1$ and $\tau_2$ as seen in Eq.~(14) of the main text.
Figure.~\ref{fig:etaC_power_emp}(a) shows the $\tau_2$-dependent behavior of $\eta_\textrm{C}^\textrm{a}$ for various values of $\tau_1$ = $0, 0.2, 0.4$, and $0.6$ with parameters $k = 2$, $\gamma_1 = 1$, $\gamma_2 = 1$, $T_1^\textrm{e}= 2$, $T_2^\textrm{e}= 1$, and $K = 2$. As the figure shows, $\eta_\textrm{C}^\textrm{a}$ is a monotonically increasing function of $\tau_2$ and gets lowered as $\tau_1$ increases.
With $\tau_1=0$ (discussed in the main text), $\eta_\textrm{C}^\textrm{a}$ is always larger than  $\eta_\textrm{C}^\textrm{e}$, whereas it becomes smaller than $\eta_\textrm{C}^\textrm{e}$ for small $\tau_2$ when $\tau_1 > 0$.
We note that a similar dependence of $\eta_\textrm{C}^\textrm{a}$ on $\tau_1$ and $\tau_2$ arises when $T_i=D_i/\gamma_i$ is fixed instead of fixed $T_i^\textrm{e}$. In the case when the heat flow is reversed, discussed in SM I.C, the efficiency should be redefined which is not shown here.

The local maximum power is obtained at the optimal value $\delta_m = \sqrt{T_2^\textrm{e} \mathcal{A}_K/(T_1^\textrm{e} \mathcal{B}_K)} \epsilon_m$ for fixed $K$. At the optimal value, the local maximum power and the efficiency at the local maximum power are given by
\begin{equation}
    \langle \dot{W} \rangle^\textrm{m}_\textrm{s}
    = \frac{K}{k(\gamma_1 + \gamma_2)} \left (
    \sqrt{\frac{T_1^\textrm{e}}{\mathcal{A}_K}} -
    \sqrt{\frac{T_2^\textrm{e}}{\mathcal{B}_K}}
    \right )^2~~\quad\textrm{and}\quad~~
    \eta_\textrm{m} = 1 - \sqrt{\frac{
    T_2^\textrm{e} \mathcal{A}_K}{
    T_1^\textrm{e} \mathcal{B}_K}},
\end{equation}
respectively. Their behaviors are similar to those of $\eta_\textrm{C}^\textrm{a}$ as shown in Figs.~\ref{fig:etaC_power_emp}(b) and (c).


\section{entropy production rate for an active-reservoir engine}

We consider an underdamped dynamics with a  velocity-dependent force with multiple equilibrium reservoirs such as
\begin{align}
d{\textbf{x}} &=\textbf{v}dt,~~\textsf{M}d{\textbf{v}} = \textbf{f}(\textbf{x},\textbf{v})dt -\textsf{G}\textbf{v}dt
+d\Xi~, \label{eqS:underdamped_Langevin}
\end{align}
where $\textsf{M}$ is a mass  matrix, $G$ is a friction matrix, and the noise satisfies
$\langle d\Xi (t)\rangle =0$ and $\langle d\Xi(t) d\Xi^{\textsf{T}} (t)   \rangle = 2\textsf{D} dt$
with $\textsf{D}=\textsf{D}^{\textsf{T}}$.
The force $\textbf{f}$ can be divided into the reversible and irreversible parts,
$\textbf{f}=\textbf{f}^{r}+\textbf{f}^{ir}$, such that
$\textbf{f}^{r}(\textbf{z})=\textbf{f}^{r}(\varepsilon\textbf{z})$
and $\textbf{f}^{ir}(\textbf{z})=-\textbf{f}^{ir}(\varepsilon\textbf{z})$, with the state vector
$\textbf{z}=(\textbf{x},\textbf{v})$ and the parity operator $\varepsilon$ defined by $\varepsilon\textbf{z}=(\textbf{x},-\textbf{v})$~\cite{Risken,Kwon}.

The environmental entropy production (EP) during an infinitesimal time increment is given by
\begin{align}
dS_\textrm{env}=\ln \frac{{\Pi}(\textbf{z}^\prime, t+dt|\textbf{z},t)}{{\Pi}(\varepsilon\textbf{z}, t+dt|\varepsilon\textbf{z}^\prime,t)}~,
\end{align}
where ${\Pi}$ is the conditional probability for a given trajectory~\cite{Kwon,ParkJKPS}. Then, the straightforward algebra yields
\begin{align}
dS_\textrm{env}=-(\textsf{M}d\textbf{v}-\textbf{f}^r dt)^\textsf{T} \textsf{D}^{-1}\circ (\textsf{G}\textbf{v}-\textbf{f}^{ir})
-\partial_{\textbf{v}} \textsf{M}^{-1}\textbf{f}^r dt~.
\end{align}
The similar expression is shown in Eq.~(11) in the reference~\cite{Kwon} for the mass matrix $\textsf{M}=m\textsf{I}$ with
the identity matrix $I$.

Then, the standard stochastic calculus gives the average environmental EP rate as
\begin{align}
\langle \dot{S}_\textrm{env} \rangle = \langle (\textbf{f}^{ir}-\textsf{G}\textbf{v})^\textsf{T} \textsf{D}^{-1}(\textbf{f}^{ir}-\textsf{G}\textbf{v}) \rangle -\textrm{Tr} (\textsf{G}\textsf{M}^{-1}) - \langle \partial_{\textbf{v}} \textsf{M}^{-1}(\textbf{f}^r-\textbf{f}^{ir})\rangle~, \label{eqS:EP_gen}
\end{align}
which is usually divided into the conventional Clausius EP term and the so-called {\em unconventional} extra EP term, see Eqs.~(25) and (26) in~\cite{Kwon,Lee_old}.
However, in the active-reservoir problem, the heat current  flowing out of the active reservoir contains the work done by the velocity-dependent force in the underdamped representation (see the main text below Eq.~(16)). Thus, it is not useful to divide the EP term in that way.

Now, we restrict ourselves to a simple linear force such as
\begin{align}
\mathbf{f}(\mathbf{x},\mathbf{v})=-\mathsf{K}\mathbf{x}+\mathsf{B}\mathbf{v}~,
\end{align}
which corresponds to our case in Eq.~(16) with $\mathbf{f}^r=-\mathsf{K}\mathbf{x}$ and  $\mathbf{f}^{ir}=\mathsf{B}\mathbf{v}$. In this case, it is convenient to rewrite the equations of motion as
\begin{align}
d{\textbf{x}} &=\textbf{v}dt,~~\textsf{M}d{\textbf{v}} =
-\mathsf{K}\mathbf{x} dt  -\tilde{\textsf{G}}\textbf{v}dt
+d\Xi~, \label{eqS:mod_Langevin}
\end{align}
with the new friction matrix $\tilde{\textsf{G}}=\textsf{G}-\mathsf{B}$, which is not diagonal with off-diagonal elements in $\textsf{B}$ for the Lorentz-like force. As the irreversible part of the force is absorbed into the friction matrix, we have the reversible force only. Then,
Eq.~\eqref{eqS:EP_gen} reads
\begin{align}
\langle \dot{S}_\textrm{env} \rangle &= \langle (\tilde{\textsf{G}}\textbf{v})^\textsf{T} \textsf{D}^{-1}(\tilde{\textsf{G}}\textbf{v}) \rangle -\textrm{Tr} (\tilde{\textsf{G}}\textsf{M}^{-1})~,\nonumber\\
 &=  \textrm{Tr}\left(
\tilde{\textsf{G}}^\textsf{T}\textsf{D}^{-1}\tilde{\textsf{G}}\Sigma^{\textbf{v}}\right)-
\textrm{Tr}\left(\tilde{\textsf{G}}\textsf{M}^{-1}  \right)~,
\label{eqS:EP_gen1}
\end{align}
where the velocity-velocity correlation matrix $\Sigma^{\textbf{v}}= \langle \textbf{v}\textbf{v}^\textsf{T}\rangle$.

It is interesting to introduce the {\em active} heat matrix differential as
\begin{align}
d\textsf{Q}^a\equiv\left[-\tilde{\textsf{G}}\textbf{v} dt +d\Xi\right]\circ \textbf{v}^\textsf{T}~.
\end{align}
The diagonal part of the heat matrix corresponds to the heat flow out of each active reservoir, discussed in the paragraph below Eq.~(16);
$\textsf{Q}^a_{ii}=Q^a_i$. A simple stochastic calculus yields
\begin{align}
\langle \dot{\textsf{Q}}^a\rangle^\textsf{T} =-\left\langle \textbf{v} (\tilde{\textsf{G}}\textbf{v})^\textsf{T}\right\rangle +
\textsf{M}^{-1}\textsf{D}~.
\end{align}
In order to relate the heat matrix to the environmental EP, we also introduce the {\em effective} temperature
matrix as
\begin{align}
{\cal{T}}^\textrm{e}\equiv \tilde{\textsf{G}}^{-1}\textsf{D}~.
\end{align}
Then, one can easily find
\begin{align}
\langle \dot{S}_\textrm{env} \rangle = - \textrm{Tr}\left[({\cal{T}}^\textrm{e})^{-1}  \langle \dot{\textsf{Q}}^a\rangle^\textsf{T}  \right]~,
\end{align}
which is similar to the {\em generalized} Clausius relation discussed in the reference~\cite{Kwon}.

We now focus on our specific problem given in Eq.~(16) with
\begin{align}
\textsf{M}= \left( \begin{array}{cc}
m_1 & 0 \\
0& m_2 \\
\end{array} \right)~,~~
\textsf{K}= \left( \begin{array}{cc}
k & -\epsilon \\
-\delta & k \\
\end{array} \right)~,~~
\textsf{B}= \left( \begin{array}{cc}
0 & \epsilon\tau_1 \\
\delta \tau_2& 0 \\
\end{array} \right)~,~~
\textsf{D}= \left( \begin{array}{cc}
D_1 & 0 \\
0& D_2 \\
\end{array} \right)~,
\tilde{\textsf{G}}= \left( \begin{array}{cc}
\Gamma_1 & -\epsilon\tau_1 \\
-\delta \tau_2& \Gamma_2 \\
\end{array} \right)~,
\end{align}
where $m_i=\gamma_i\tau_i$.
In the steady state, the environment EP is equivalent to the total EP, $\langle \dot{S}\rangle_\textrm{s}$, and
$\Sigma^{\textbf{v}}$ can be easily calculated from the steady-state covariant matrix for the active-reservoir engine in Eq.~\eqref{CM_active}, using the identities as
\begin{align}
\Sigma^\textbf{v}_{ij} = \langle \textbf{v}_i\textbf{v}_j\rangle_\textrm{s}=
\langle \dot{\textbf{x}}_i\dot{\textbf{x}}_j\rangle_\textrm{s}=\langle (\textsf{A}\textbf{z})_i(\textsf{A}\textbf{z})^\textsf{T}_j\rangle_\textrm{s}=
(\textsf{A}\Sigma\textsf{A}^\textsf{T})_{ij}~\quad (i,j=1,2)~,
\end{align}
where
$\textsf{A}$ and $\Sigma$ are the linear force matrix and the covariant matrix in Eq.~\eqref{CM_active}, respectively.

After a lengthy but straightforward algebra for Eq.~\eqref{eqS:EP_gen1}, we finally obtain the total EP in the steady state of the active-reservoir engine as
\begin{align}
 \langle \dot{S}\rangle_\textrm{s}=-\frac{\langle \dot{Q}_1\rangle_\textrm{s} \mathcal{A}_K}{T_1^\textrm{e}}-\frac{\langle \dot{Q}_2\rangle_\textrm{s} \mathcal{B}_K}{T_2^\textrm{e}}
 +(\tau_1-\tau_2)\left[ \frac{\tau_1 \epsilon^2}{\tau_2 \gamma_2\Gamma_1\mathcal{B}_K}\left(\frac{T_2^\textrm{e}}{T_1^\textrm{e}}\right)
 -\frac{\tau_2 \delta^2}{\tau_1  \gamma_1\Gamma_2\mathcal{A}_K}\left(\frac{T_1^\textrm{e}}{T_2^\textrm{e}}\right) \right]~, \label{eqS:EP}
\end{align}
with $\langle\dot{Q}_i\rangle_\textrm{s}=\langle\dot{Q}_i^a\rangle_\textrm{s}=\langle \dot{\textsf{Q}}_{ii}^a\rangle_\textrm{s}$, which is given by Eq.~\eqref{eqS:x1v2_active}. As usual, the thermodynamic second law guarantees the non-negativity of the total EP, i.e., $\langle \dot{S}\rangle_\textrm{s}\ge0$.
Note that the first two terms look like Clausius EP terms with the {\em modified} effective temperatures $T_1^\textrm{e}/\mathcal{A}_K$
and $T_2^\textrm{e}/\mathcal{B}_K$, but the third term seems not related to any simple physical observable.
The unconventional EP term appearing generally in the underdamped dynamics with velocity-dependent forces~\cite{Kwon}
contributes to both the modified Clausius EP term and the extra third term.
Interestingly, the modified Clausius EP term turns out to be always non-negative in this specific problem, i.e.~
\begin{align}
\langle \dot{S}_{mClau}\rangle_\textrm{s}=-\frac{\langle \dot{Q}_1\rangle_\textrm{s} \mathcal{A}_K}{T_1^\textrm{e}}-\frac{\langle \dot{Q}_2\rangle_\textrm{s} \mathcal{B}_K}{T_2^\textrm{e}}=\frac{\mathcal{A}_K \mathcal{B}_K}{k(\gamma_1+\gamma_2)T_1^\textrm{e} T_2^\textrm{e}}
\left[\frac{T_1^\textrm{e}\delta }{\mathcal{A}_K}-\frac{T_2^\textrm{e}\epsilon }{\mathcal{B}_K}\right]^2 \ge 0~.
\end{align}
Along the maximum efficiency line ($\eta=\eta_\textrm{C}^{a}$) in Fig.~1(d) derived from Eq.~(12), this modified Clausius EP rate vanishes just like in the
equilibrium-reservoir case where the maximum efficiency is obtained in the reversible process (zero EP rate).
However, the true EP  along the maximum efficiency line does not vanish due to the extra EP term in
Eq.~\eqref{eqS:EP}.

As the value of the extra EP term can be positive, negative, or zero depending on the parameter values, there is no inequality between the true EP and the modified Clausius EP, in general. Nevertheless, when the persistence time is the same for both active reservoirs ($\tau_1=\tau_2$), the modified Clausius EP becomes exactly identical to the total EP. Therefore, in this special case, all physical quantities including the heat, work,  EP for the active-reservoir engine dynamics turn out be equivalent to those for the equilibrium-reservoir engine with the modified effective temperatures, $T_1^\textrm{e}/\mathcal{A}_K$ and $T_2^\textrm{e}/\mathcal{B}_K$.
For a more special case with the symmetric time scale ($\tau_1=\tau_2$ and $\gamma_1=\gamma_2$) discussed before, the (modified effective) temperature ratio is identical to the equilibrium temperature ratio and all physical quantities become identical to
those for the equilibrium-reservoir engine with the temperatures $T_1$ and $T_2$ except for the overall time-scale factor.


\vfil\eject